\documentstyle[11pt,aaspp4,amssym,epsf,twoside]{article}

\newcommand{\pic}[4]{
\noindent\parbox{#1}{\hbox{\epsfxsize=#1\epsfbox{#2}}
\caption{\protect\small\em #3\label{#4}}
}}

\newcommand{\picw}[4]{
\noindent\parbox{\textwidth}{\hbox{\centerline{\epsfxsize=#1\epsfbox{#2}}}
\caption{\protect\small\em #3\label{#4}}
}}

\begin{document}


\title{Analyzing X-Ray Pulsar Profiles:\\
       Geometry and Beam Pattern of Her~X-1}

\author{S. Blum and U. Kraus}
\affil{Institut f\"ur Astronomie und Astrophysik,
       Abteilung Theoretische Astrophysik,\\
       Universit\"at T\"ubingen,\\
       Auf der Morgenstelle 10,
       72076 T\"ubingen, Germany}


\begin{abstract}

We report on our analysis of a large sample of energy dependent pulse
profiles of the X-ray binary pulsar Hercules X-1. We find that all data are
compatible with the assumption of a slightly distorted magnetic dipole field
as sole cause of the asymmetry of the observed pulse profiles. Further the
analysis provides evidence that the emission from both poles is equal. We
determine an angle $\Theta_{\rm m} < 20^{\circ}$ between the rotation axis
and the local magnetic axis. One pole has an offset $\delta<5^{\circ}$ from
the antipodal position of the other pole. The beam pattern shows structures
that can be interpreted as pencil- and fan-beam configurations. Since no
assumptions on the polar emission are made, the results can be compared with
various emission models. A comparison of results obtained from pulse
profiles of different phases of the 35-day cycle indicates different
attenuation of the radiation from the poles being responsible for the change
of the pulse shape during the main-on state. These results also suggest the
resolution of an ambiguity within a previous analysis of pulse profiles of
Cen X-3, leading to a unique result for the beam pattern of this pulsar as
well. The analysis of pulse profiles of the short-on state indicates that a
large fraction of the radiation cannot be attributed to the direct emission
from the poles. We give a consistent explanation of both the evolution of
the pulse profile and the spectral changes with the 35-day cycle in terms of
a warped precessing accretion disk.

\end{abstract}

\section{Introduction}
\label{intro}

Since its discovery in 1972 by the UHURU satellite (Tananbaum et al. 1972),
the X-ray binary system Hercules~X-1/HZ~Herculis has become the best studied
of its class of about 44 known today (Bildsten et al. 1997). They are
understood to be fast spinning neutron stars that are accreting matter from
a massive companion star either via Roche lobe overflow or from the stellar
wind of the companion. Since the neutron stars have strong magnetic fields,
the accreted matter is funnelled along the field lines onto the magnetic
poles, where most of the energy is released in form of X-radiation.
Generally the magnetic axis and the rotation axis are not aligned. Therefore
a large fraction of the detected flux from these sources is pulsed as during
the course of each revolution of the neutron star the beams from the poles
sweep through our line of sight.

Her X-1/HZ Her combines most of the properties that can be found in X-ray
binaries, this made it one of the favourite sources of X-ray astronomers.
From the observation of eclipses and from pulse timing analyses the orbital
parameters are well determined. The masses of the neutron star and its
optical companion are $1.3 \ M_{\odot}$ and $2.2 \ M_{\odot}$ respectively,
the orbital period of Her X-1 is $1.7 \ {\rm d}$, and the inclination of the
orbital plane is $i>80^{\circ}$ (Deeter, Boynton, \& Pravdo 1981). In
addition to the pulse period of 1.24 s, i.e. the rotation period of the
neutron star, Her X-1 also displays X-ray intensity variations on a period
of about 35 days. Such a long-term variability is only known for two other
pulsars: LMC~X-4 and SMC~X-1. The 35-day cycle of Her~X-1 is nowadays
ascribed to the precession of a warped accretion disk which periodically
obscures the neutron star from our view (Petterson, Rothschild, \& Gruber
1991, Schandl \& Meyer 1994). During its high intensity or main-on state,
Her X-1 has a luminosity $L_{\rm x} \approx 2.5 \cdot 10^{37} \ {\rm ergs \
s}^{-1}$ (2-60~keV) (McCray et al. 1982). The maximum flux of the short-on
state is typically only 30\% of that of the main-on.  Balloon observations
in 1977 allowed for the first time the indirect measurement of the magnetic
field strength of some $10^{12}$ G by the revelation of a spectral feature
in the hard X-ray spectrum (Tr\"umper et al. 1978), interpreted as a
cyclotron absorption line at about 40 keV.

The pulse shapes of Her~X-1 are highly asymmetric and depend on energy and
on the phase of the 35-day cycle. In several studies phenomenological
emission patterns have been used to reproduce the asymmetric pulse profiles
of Her~X-1. Wang \& Welter (1981) fitted the geometry of two antipodal polar
caps with asymmetric fan-beam patterns. In this approach the asymmetry of
the emission pattern was attributed to asymmetric accretion due to the
plasma becoming attached to the magnetic field lines away from the
corotation radius. However it is not clear whether an asymmetric accretion
stream must produce an asymmetric beam pattern (Basko \& Sunyaev
1975). Another way of introducing asymmetry into the pulse shapes is via non
antipodal emission regions. Leahy (1991) used two offset rings on the
surface of the neutron star with symmetric pencil-beams and Panchenko \&
Postnov (1994) modelled two antipodal polar caps and one ringlike area which
was attributed to a non-coaxial quadrupole configuration of the magnetic
field.  Further studies have shown that relativistic light deflection near
the neutron star plays an important role when emission models are used to
explain the observed pulse shapes (e.g. Riffert et al. 1993, Leahy~\&~Li
1995).

In this analysis we take up the idea of a non antipodal location of the
emission regions caused by a slightly distorted magnetic dipole field. We
assume that the emission originating from the regions near the magnetic
poles only depends on the viewing angle between the magnetic axis and the
direction of observation which means that the emission is symmetric with
respect to the local magnetic axis. In contrast to previous studies where
specific emission models have been used to fit the pulse profiles, the
method used here does not involve any assumptions on the polar emission.
Instead it tests in a general way whether the pulse profiles are compatible
with the assumption that they are the sum of two independent symmetric
components.

The method we use to analyze pulse profiles is briefly summarized in the
following \S\ref{method}. In \S\ref{data} we list the analyzed data. The
results of the analysis are presented in \S\ref{results}. We show that the
data of Her~X-1 are indeed compatible with the idea of a slightly distorted
magnetic dipole field. Further we find indications in the contributions to
the pulse profiles that the emission from both poles is identical. We
determine the location of the magnetic poles and reconstruct the beam
pattern, which is discussed in \S\ref{char}. In the following \S\ref{35d} we
examine the dependence of the pulse shape on the phase of the 35-day cycle.
We argue that the contributions to the pulse profile undergo different
attenuation resulting in the observed evolution of the pulse shapes during
the main-on state of the 35-day cycle.

\section{Analysis}
\label{analysis}

\subsection{The Method}
\label{method}

This section is a short summary of the method we use to analyze the energy
dependent pulse profiles of Her~X-1. We will focus on the main ideas and
assumptions omitting both formal derivations and technical details. A
comprehensive presentation of the material including a test case has been
given in Kraus et al. (1995).

Consider the emission region near one of the magnetic poles of the neutron
star. Radiation escapes from the accretion stream and from the star's
surface and, while close to the star, is deflected in the gravitational
field of the neutron star. A distant observer who cannot spatially resolve
the emission region measures the integrated flux coming from the entire
visible part of the emission region. The observed integrated flux depends on
the direction of observation because the direction of observation determines
which part of the emission region is visible and also because the radiation
emitted by the accretion stream and the neutron star is presumably
beamed. This function, namely the flux of a single emission region measured
by a distant observer as a function of the direction of observation, is the
link between the properties of the emission region and the contribution of
that emission region to the pulse profile. In the following we will call
this function the beam pattern of the emission region. The contribution of
the emission region to the pulse profile, which we will refer to as a
single-pole pulse profile, depends both on the beam pattern and on the
pulsar geometry, i.e., on the orientation of the rotation axis with respect
to the direction of observation and on the location of the magnetic pole on
the neutron star. In short: local emission pattern plus relativistic light
deflection determine the beam pattern, beam pattern plus geometry result in
a certain single-pole pulse profile and the superposition of the single-pole
pulse profiles of the both emission regions is the total pulse profile.

\subsubsection*{a. decomposition into single-pole pulse profiles}

In the following we are going to assume that the beam pattern is
axisymmetric with respect to the magnetic axis (i.e., to the axis that
passes through the center of the neutron star and through the magnetic
pole). The axisymmetric beam pattern is a function of only one variable, the
angle $\theta$ between the direction of observation and the magnetic axis.
Consider now the single-pole pulse profile $f(\phi)$, where $\phi$ is the
angle of rotation of the neutron star. It can easily be shown that the
single-pole pulse profile produced by an axisymmetric beam pattern is
symmetric in the following sense: there is a rotation angle $\Phi$, so that
$f(\Phi-\phi) = f(\Phi+\phi)$ for all values of $\phi$. The fact that $f$ is
periodic in $\phi$ implies that the same symmetry must hold with respect to
the rotation angle $\Phi+\pi$.

Now turn to the total pulse profile produced as the sum of the two symmetric
single-pole pulse profiles. If the emission regions are antipodal, i.e., the
two magnetic axes are aligned, it turns out that the symmetry points
$\Phi_1$ and $\Phi_1 + \pi$ of the first single-pole pulse profile fall on
the same rotation angles as the symmetry points $\Phi_2$ and $\Phi_2 + \pi$
of the second single-pole pulse profile. Their sum, the total pulse profile,
is therefore symmetric with respect to the same symmetry points. If the
emission regions are not antipodal, however, the symmetry points of the two
single-pole pulse profiles do not coincide (except for certain special
displacements from the antipodal positions) and the total pulse profile is
asymmetric.

Given an observed asymmetric pulse profile, we can ask if it could possibly
have been built up out of two symmetric contributions with symmetry points
that do not coincide. If so, it must be possible to find two symmetric (and
periodic) functions $f_1$ and $f_2$ with the pulse profile $f$ as their sum.
By writing the observed pulse profile, defined by a certain number $N$ of
discrete data points $f(\phi_k)$, as a Fourier sum and with an ansatz for
$f_1$ and $f_2$ in the form of Fourier sums also, the following can easily
be shown: For an arbitrary choice of symmetry points $\Phi_1$ and $\Phi_2$,
there are two periodic functions $f_1$ and $f_2$, $f_1$ symmetric with
respect to $\Phi_1$ and $f_2$ symmetric with respect to $\Phi_2$, such that
$f=f_1 + f_2$, and the two symmetric functions are uniquely determined.
Exceptions to this rule occur only if $(\Phi_1-\Phi_2)/\pi$ is a rational
number. In this case the symmetric functions may not exist or, if they
exist, may not be uniquely determined. It must also be noted that the
symmetric functions obviously can only be determined up to a constant $C$,
since $f_1 + C$ and $f_2 - C$ are also a solution if $f_1$ and $f_2$ are.

Thus, in principle every choice of a pair of symmetry points corresponds to
a unique decomposition of any pulse profile into two symmetric
contributions. For such a decomposition to be an acceptable solution,
however, $f_1$ and $f_2$ also have to meet the following physical criteria
in order to be interpreted as single-pole pulse profiles:
\begin{enumerate}
\item They must not have negative values, since they represent photon fluxes.
\item They must be reasonably simple and smooth.
      We do not expect the polar contributons to have a shape that is more
      complex than the pulse profile. Especially modulations of the
      single-pole pulse profiles that cancel out in the sum are not
      compatible with the assumption of two independent and therefore
      uncorrelated emission regions.
\item They must conform to the energy dependence of the pulse profile.
      The decomposition can be done independently for pulse profiles in
      different energy ranges. Since the symmetry points are determined by
      the pulsar geometry, the same symmetry points must give acceptable
      decompositions according to the criteria 1 and 2 in all energy ranges.
      Finally the single-pole pulse profiles should show the same gradual
      energy dependence as the pulse profile.
\end{enumerate}

Given the existence of formal decompositions for all pairs of symmetry
points and the criteria mentioned above, we are left with the
two-dimensional parameter space of all possible values of $\Phi_1$ and
$\Phi_2$, which we search for points with acceptable decompositions. For
practical purposes, the parameters we use are the quantities $\Phi_1$ and
$\Delta := \pi - (\Phi_1-\Phi_2)$. The parameter space that contains every
possible unique decomposition then is $0 \leq \Phi_1 \leq \pi$ and $0 \leq
\Delta \leq \pi/2$. In the analysis of just one pulse profile there will in
general be a number of different acceptable decompositions. This number may
be significantly reduced by the energy dependence of the pulse profile. In
general, the existence of symmetry points with acceptable decompositions in
all energy channels is by no means guaranteed. If such a pair of symmetry
points is found, then it is indeed possible to build up the observed pulse
profile out of two symmetric contributions, and we can conclude that the
analyzed data are compatible with the assumption that the asymmetry of the
observed pulse profile is caused by the non-antipodal locations of the
magnetic poles. The symmetric functions can be interpreted as the
single-pole pulse profiles due to the two emission regions.

A successful decomposition provides information both on the geometry and on
the beam pattern. As to the geometry, we obtain a value for the parameter
$\Delta$. This parameter is related to the locations of the emission regions
on the neutron star (see Figure~\ref{fig1}). The beam pattern is related to
the single-pole pulse profile via the geometric parameters, i.e., the
location of the emission region on the neutron star and the direction of
observation. Since these parameters are not known, one cannot directly
deduce the beam pattern from the single-pole pulse profile. It can be shown,
however, that an appropriate transformation of the single-pole pulse profile
and the beam pattern turns the transformed single-pole pulse profile into a
scaled, but undistorted copy of a section of the transformed beam pattern.
Although the scaling factor is a geometric quantity and therefore not known,
this still provides an intuitive understanding of what a section of the beam
pattern must look like. Since in the case of Her~X-1 it is possible to
eventually reconstruct the beam pattern, the information obtained at this
stage mainly serves as a starting point for the next step of the analysis
and we will not go into details about the transformation mentioned above.

\begin{figure}
\picw{3.0in}{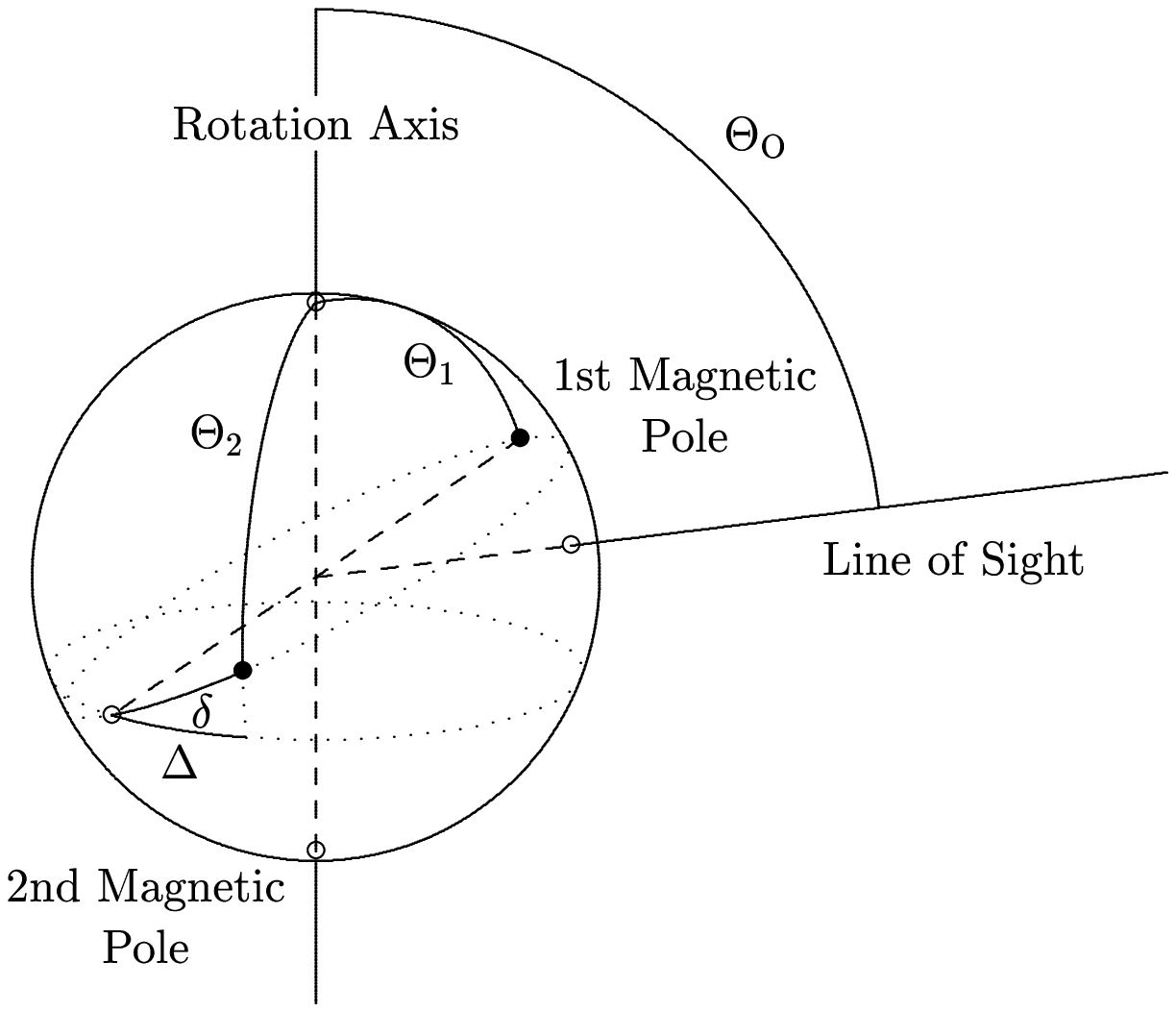}{%
Intrinsic pulsar geometry: the locations of the magnetic poles on the
neutron star surface can be described by means of their polar angles
$\Theta_1$ and $\Theta_2$ with respect to the rotation axis and by the
angular distance $\delta$ between the location of one magnetic pole and the
point that is antipodal to the other magnetic pole.
}{fig1}
\end{figure}

\subsubsection*{b.
search for an overlap region and determination of the geometry}

In general, the two emission regions on the neutron star may or may not be
equal (i.e., have the same beam pattern). If they are equal, this fact may
be apparent in the single-pole pulse profiles in the following way. Since in
general the rotation axis and the magnetic axis of the neutron star are not
aligned, the viewing angle $\theta$ between the magnetic axis and the
direction of observation of each emission region changes with rotation angle
$\phi$. The range $\theta$ can cover for each magnetic pole depends on the
location of that pole on the neutron star and on the direction of
observation, where $0^\circ \leq \theta_{\rm min} \leq \theta_{\rm max} \leq
180^\circ$. Only in the special case where both the magnetic axis and the
direction of observation are perpendicular to the rotation axis, $\theta$
takes all values between $0^\circ$ and $180^\circ$. Since the emission
regions have different locations on the neutron star, their ranges of values
of $\theta$ are different. Depending on the geometry, these two ranges for
$\theta$ may overlap. For an ideal dipole configuration e.g., the condition
under which an overlap in the ranges of values of $\theta$ of the both poles
exists is $\Theta_{\rm O}+\Theta_{\rm m}>\pi/2$, where $\Theta_{\rm O}$ is
the angle between the rotation axis and the line of sight, and $\Theta_{\rm
m}$ is the angle between the rotation axis and the magnetic axis. Consider
an angle $\tilde{\theta}$ in the overlap region. At some instant during the
course of one revolution of the neutron, at rotation angle $\phi$, one
emission region is seen under the angle $\tilde{\theta}$. At a different
instant, at rotation angle $\phi'$, the other emission region is seen under
the same angle $\tilde{\theta}$. If the beam patterns of the two emission
regions are identical, then the flux detected from the one emission region
at $\phi$ is equal to the flux detected from the other emission region at
$\phi'$. Thus, if an overlap region exists, the corresponding part of the
beam pattern shows up in both single-pole pulse profiles, though at
different values of rotation angle.  Since the single-pole pulse profiles
can be transformed into undistorted (though scaled) copies of sections of
the beam patterns, such a part of the beam pattern that shows up in both
single-pole pulse profiles should be readily recognizable. Note that the
occurence and size of the overlap region depends on the geometric parameters
and must therefore be the same for pulse profiles in different energy
channels.

If an overlap region is found in the single-pole pulse profiles obtained in
the decomposition, this is an indication that there are two emission regions
with identical beam patterns. Since each single-pole pulse profile provides
a section of the beam pattern and the two sections overlap, we can then
combine the two sections by superposing the overlapping parts. As a result
we obtain the total visible section of the beam pattern. Superposing the
overlapping parts of the two sections of the beam pattern amounts to
determining the relation between the corresponding values $\phi$ and $\phi'$
of the rotation angle. On the other hand, the relation between $\phi$ and
$\phi'$ can be expressed in terms of the unknown geometric parameters of the
system. Thus, the superposition provides a constraint on the geometry.

Again omitting all details we simply note the procedure for superposing the
overlapping parts of the two sections of the beam pattern. The single-pole
pulse profiles $f_1(\phi)$ with symmetry point $\Phi_1$ and $f_2(\phi)$ with
symmetry point $\Phi_2$ are transformed into functions of a common variable
$q$ through $\cos(\phi-\Phi_1) = q$ for $f_1$ and
$\cos(\phi-\Phi_2)=(q-a)/b$ for $f_2$. The real numbers $a$ and $b>0$ are
determined by means of a fit which minimizes the quadratic deviation between
$f_1(q)$ and $f_2(q)$ in the overlap region. At this point the constant $C$,
which determines how the unpulsed flux has to be distributed to the
single-pole pulse profiles, can also be computed. Since $a$ and $b$ can be
expressed in terms of the unknown geometric parameters of the pulsar, their
best-fit values constitute constraints on the pulsar geometry. The results
of this second step of the analysis are the total visible beam pattern as a
function of $q$ and two constraints on the geometric parameters.

The geometric information obtained so far (i.e., the values of $\Delta$,
$a$, and $b$) is not quite sufficient in itself to completely determine the
pulsar geometry. It needs to be supplemented by an independent determination
of any one additional geometric parameter or by an additional constraint. We
suggest that this supplement may be obtained by means of the assumption that
the rotation axis of the neutron star is perpendicular to the orbital
plane. In this case, the angle $\Theta_{\rm O}$ between the direction of
observation and the rotation axis of the neutron star is given by the
inclination of the orbital plane. The assumption of $\Theta_{\rm O}=i$ seems
to be quite plausible since accreted mass also carries angular momentum from
the massive companion, and this transfer is expected to align the rotation
axes of the binary stars on a timescale short compared to the lifetime of
the system. However, this assumption must not hold true for all binary
systems. With the inclination substituted for $\Theta_{\rm O}$, the analysis
of the pulse profiles determines the positions of the emission regions on
the neutron star.

Once the pulsar geometry is known, we also obtain the equation relating the
auxiliary variable $q$ and the viewing angle $\theta$, so that the
reconstructed beam pattern can be transformed into a function of $\theta$.
However, it turns out that the relation between $q$ and $\theta$ involves an
ambiguity which cannot be resolved within this analysis. It is due to the
fact that we are not able to relate a single-pole pulse profile to one of
the two emission regions. Therefore, we obtain two different possible
solutions for the beam pattern and a choice between them must be based on
either theoretical considerations and model calculations, or on additional
information on the source.

\subsection{The Data}
\label{data}

The analysis presented in this paper is based on pulse profiles of the
main-on and short-on states of Her~X-1. The analyzed sample contains a total
of 148 pulse profiles from 20 different observations. References, the
platform of the detectors, year of observation, the total energy range, the
state of the 35-day cycle, the number of separate observations and the total
number of pulse profiles of the respective observations are listed in
Table~\ref{tab1}. The data reduction including background subtraction has
been done by the respective authors. In order to compare the pulse profiles
from different observations, the pulse profiles of the main-on have been
aligned in phase so that their common features match best. Since the pulse
profiles of the short-on are markedly different, their features have been
aligned with respect to the main-on as suggested by Deeter et al. (1998).

\begin{deluxetable}{lccrccc}
\tablecaption{Analyzed Data \label{tab1}}
\tablewidth{0pt}
\tablehead{
\colhead{} & \colhead{Instr.} &
\colhead{Year of} & \colhead{Energy Range} &
\colhead{35-Day} & \colhead{No. of} & \colhead{Pulse} \\
\colhead{Reference} & \colhead{Platform} & \colhead{Obs.} & \colhead{(keV)} &
\colhead{State} & \colhead{Obs.\tablenotemark{a}} &
\colhead{Profiles\tablenotemark{b}}}
\startdata
Kuster et al. 1998 & RXTE & 1997 & 2 - 19~~~~~~~ & turn-on & 1 & 25 \nl
Soong et al. 1990a & HEAO-1 & 1978 & 12 - 55~~~~~~~ & main-on &
1\tablenotemark{c} & ~5 \nl
Kahabka 1987 & EXOSAT & 1984/1985 & 0.9 - 29~~~~~~~ & main-on & 4 & 48 \nl
Kunz 1996\tablenotemark{d} & Kvant & 1987/1988 & 16 - 30~~~~~~~ & main-on &
1\tablenotemark{c} & ~1 \nl
Scott 1993 & Ginga & 1988-1990 & 1 - 37~~~~~~~ & main-on & 8 & 40 \nl
Stelzer 1997\tablenotemark{d} & RXTE & 1996 & 2 - 26~~~~~~~ & main-on & 1 &
~9 \nl
Kahabka 1987 & EXOSAT & 1984 & 0.9 - 23~~~~~~~ & short-on & 1 & 11 \nl
Scott 1993 & Ginga & 1989 & 1 - 14~~~~~~~ & short-on & 3 & ~9 \nl

\enddata

\tablenotetext{a}{Number of separate observations}
\tablenotetext{b}{Total number of pulse profiles in different energy subranges}
\tablenotetext{c}{Several pointings have been integrated}
\tablenotetext{d}{private communication}

\end{deluxetable}

At energies below 1 keV, the pulses of Her X-1 have a sinusoidal shape which
is interpreted as reprocessed hard X-radiation at the inner edge of the
accretion disk (McCray et al. 1982). Since the origin of these soft X-rays
is not the region near the magnetic poles, the analysis is restricted to
higher energies. Above 1 keV the pulse profiles of Her X-1 are highly
asymmetric and their typical energy dependence has been examined in a
variety of studies (see Deeter et al. 1998, and references therein). In the
analysis the pulse profiles are written as Fourier series. Since the higher
Fourier coefficients are presumably affected by aliasing and also may have
fairly large statistical errors, the highest coefficients are set to zero.
This has a smoothing effect depending on the number of Fourier coefficients
concerned. An example of the typical energy dependence of the pulse profiles
and their representation in the analysis is given in the top row of
Figure~\ref{fig2}. It shows pulse profiles in three different energy ranges
of an EXOSAT observation (Kahabka 1987) during the main-on state. The
observed pulse profiles are plotted with crosses. The profiles plotted as
solid lines are inverse Fourier-transformed using 32 out of originally 60
Fourier coefficients.

\begin{figure*}\begin{center}
\pic{6.5in}{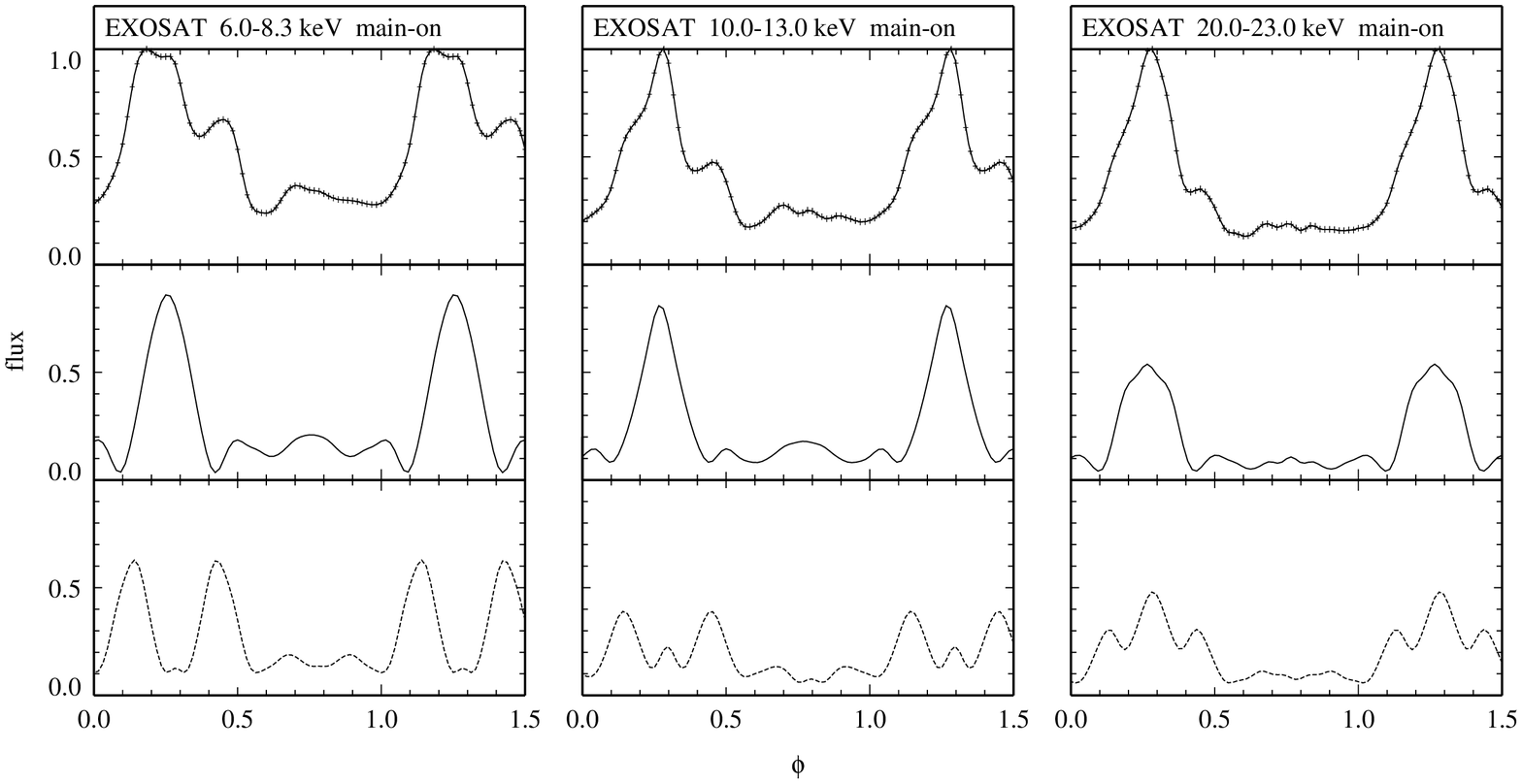}{%
Examples of analyzed pulse profiles and their decompositions into
single-pole contributions. The panels in the top row show observed pulse
profiles:
the crosses correspond to the bin-width (60 phase bins) and the estimated
statistical error of the observations, whereas the solid curves are the
inverse Fourier-transformed pulse profiles using 32 Fourier
coefficients. The pulse profiles have been normalized to have a maximum flux
of unity. The energy range is indicated above each column.
The energy dependence is typical for pulse profiles of the main-on state.
The middle and bottom panels show the decompositions of the pulse profiles
into the two symmetric contributions from the poles with symmetry points at
$\Phi_1\approx95^\circ$ ($\phi \approx 0.26$) and $\Phi_1+\Delta$ with
$\Delta\approx9^\circ$ ($\phi \approx 0.29$). The single-pole pulse profiles
add up exactly to the pulse profile in the respective top panel.
}{fig2}
\end{center}\end{figure*}

\subsection{Results}
\label{results}

\subsubsection*{a. decomposition into single-pole pulse profiles}

In a first run, the decomposition method has been simultaneously applied to
the 103 pulse profiles of the 15 observations of the main-on state. Due to
the large number of distinct pulse shapes and due to the fact that they have
a relatively low level of unpulsed flux, the positive flux criterion has led
to an exclusion of about 90\% of the whole parameter space of possible
symmetry points $\Phi_1$ and $\Phi_1+\Delta$. Further sorting out the
decompositions (i.e. the single-pole pulse profiles) that are qualitatively
too complicated to match the criterion of two independent emission regions
only left over one type of decomposition. The energy dependence of this type
of decomposition is as smooth as that of the pulse profiles. Thus we have
found acceptable decompositions in a small range of $\Phi_1$ and
$\Phi_1+\Delta$ which are all of the same type. This type of decomposition
is unique in the sense that a small deviation from the 'best-values' of the
symmetry points results in decompositions that look similar but become more
and more complicated the larger the deviation becomes until they do not
match the physical criteria any more. A systematic variation of the
best-values of the symmetry points, which could be caused by free precession
of the neutron star, is not observed. The lower panels in Figure~\ref{fig2}
show the decompositions of the typical pulse profiles of the respective top
panels. The unpulsed flux has been distributed to the single-pole pulse
profiles according to the constant $C$ as derived in the second step of the
analysis (see \S~\ref{method}). The single-pole pulse profiles show that the
energy dependence of the pulse profiles is mainly due to the change of one
polar contribution (dashed curve) where an additional peak appears above 10
keV, whereas the pulse shape of the other pole (solid curve) does not change
much. Interestingly, the contributions of the emission regions we obtain
look very similar to those of Panchenko \& Postnov (1994) obtained from a
model calculation mentioned in \S~\ref{intro}. Similar components were also
obtained by Kahabka (1987) in an attempt to model the observed pulse shapes
by means of 3 to 5 gaussians, a sinusoidal component and a constant flux.

Extending the analysis to the short-on and the turn-on of the main-on we
also find acceptable decompositions in the same range of the symmetry points
as in the main-on. Since the pulses of the short-on state have quite a
different shape compared to the main-on, their decompositions look different
as well. An example of a typical short-on pulse profile and its
decomposition is given in Figure~\ref{fig3}.

\begin{figure}
\picw{2.16in}{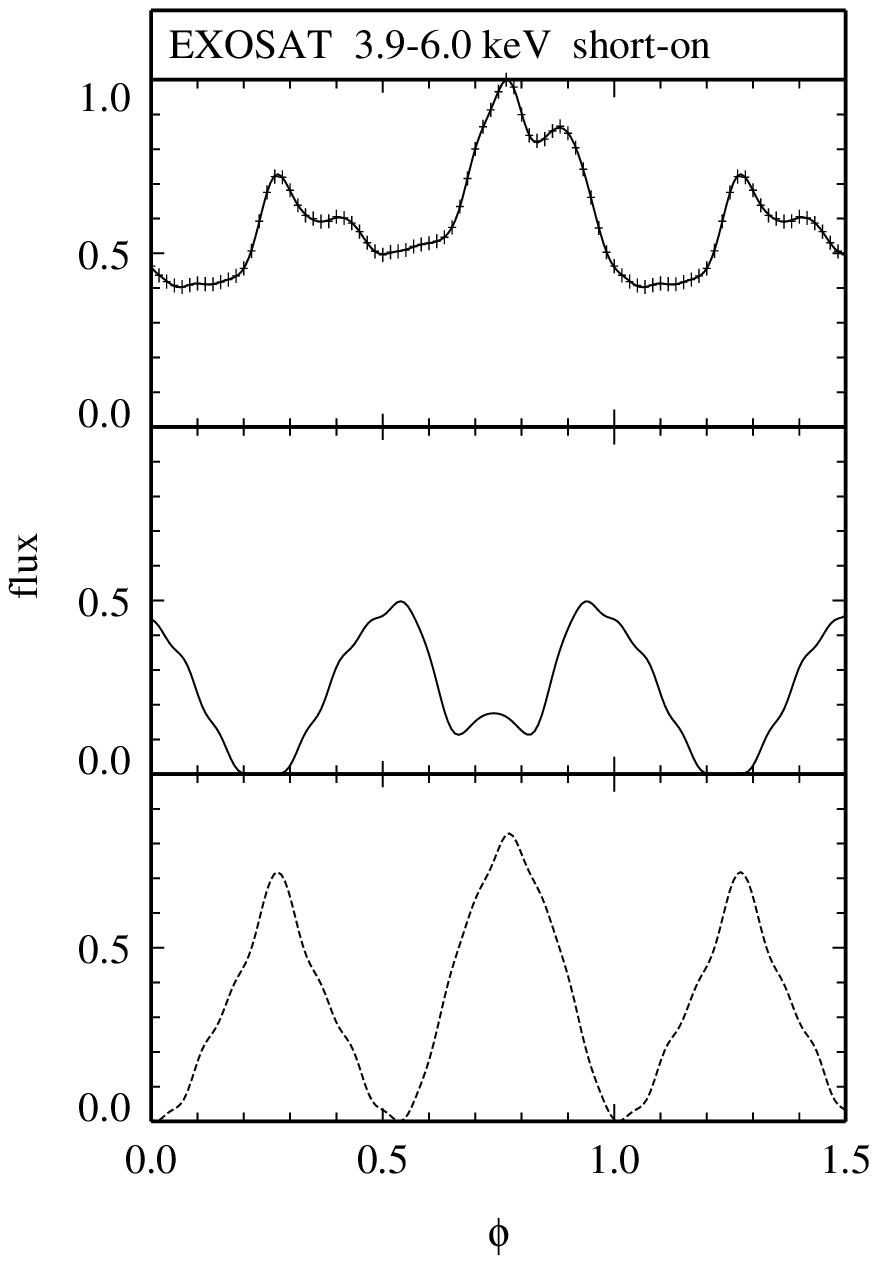}{%
Pulse profile of an EXOSAT observation (Kahabka 1987) during the short-on
state and its decomposition. Plot symbols and normalization are as in
Figure~\ref{fig2}.
}{fig3}
\end{figure}

\subsubsection*{b.
search for an overlap region and determination of the geometry}

In the next step of the analysis a two parameter fit has been applied to the
decompositions in order to find out whether there is a range where the
shapes of the polar contributions match. For the 62 pulse profiles of 8 of
the main-on observations we have found a set of fit parameters which
correspond to an overlap range where the two curves of each decomposition
match astonishingly well if the statistical errors of the data are taken
into account. This is shown in Figure~\ref{fig4} where the solid (dashed)
curve corresponds to the respective single-pole pulse profile in
Figure~\ref{fig2}. The typical errors ($\pm\sigma$) as derived from error
propagation of the statistical errors of the data are indicated in the upper
right corner of each panel. The range where the curves overlap corresponds
to values of the viewing angle $\theta$ under which both emission regions
are seen during the course of one revolution of the neutron
star. Introducing a scaling factor as an additional fit parameter we have
achieved acceptable fits for the 15 pulse profiles of another three main-on
observations. These profiles are further discussed in \S\ref{35d}. No
acceptable fits have been achieved for only four observations at late phases
of the main-on when the flux had already dropped to less than 60\% of the
maximum flux of the respective 35-day cycle.

The dependence of the results for the location of the magnetic poles on the
direction of observation $\Theta_{\rm O}$ is shown in Figure~\ref{fig5}.
Assuming that $\Theta_{\rm O}$ is equal to the inclination $i$ of the system
and adopting $i=83^{\circ} (\pm4^{\circ})$ (Kunz 1996, private
communication), we obtain the polar angles of the magnetic poles $\Theta_1
\approx 18^{\circ}$ and $\Theta_2 \approx 159^{\circ}$ with an offset from
antipodal positions of $\delta < 5^{\circ}$ (see Figure~\ref{fig1}). The
small value obtained for $\delta$ confirms the assumption that a fairly
small distortion of the magnetic dipole field is enough to explain the
considerable asymmetry of the pulse profiles of Her X-1. The error bars at
$\Theta_{\rm O}=83^{\circ}\pm4^{\circ}$ demonstrate how little the best
fit-parameters determined for different pulse profiles vary.

\begin{figure*}\begin{center}
\pic{6.5in}{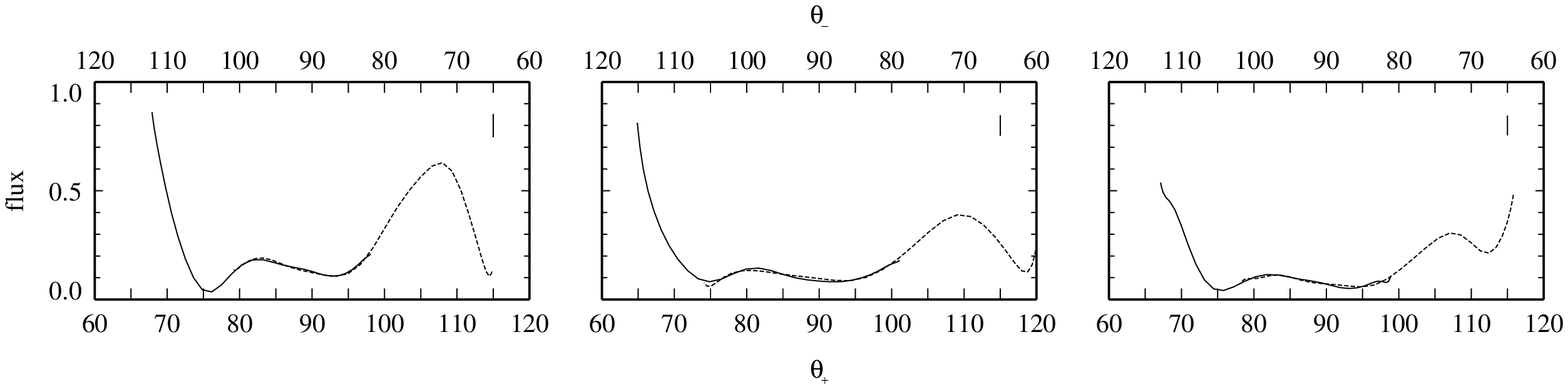}{%
Reconstructed beam patterns corresponding to the pulse profiles and their
decompositions in Figure~\ref{fig2} as functions of the viewing angle
$\theta$ between the local magnetic axis and the direction of observation.
The $\theta_+$- and the $\theta_-$-solutions correspond to the lower and
upper x-axis respectively. The typical errors ($\pm\sigma$) are indicated in
the upper right corner of each panel. The scale of the x-axis corresponds to
$\Theta_{\rm O}=83^{\circ}$.
}{fig4}
\end{center}\end{figure*}

\begin{figure}
\picw{3.0in}{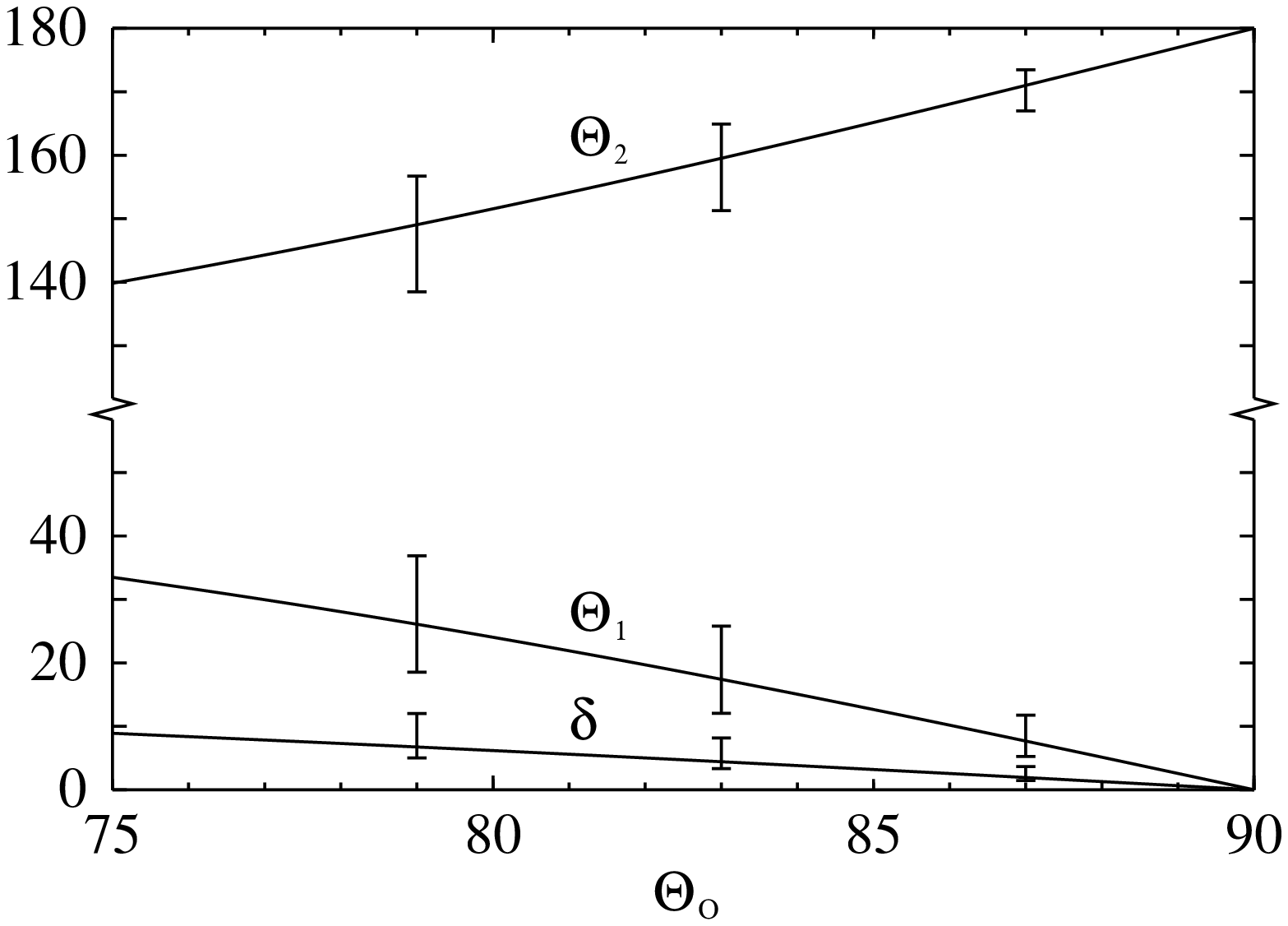}{%
Location of the magnetic poles in terms of their polar angles $\Theta_1$,
$\Theta_2$ and their offset $\delta$ as described in Figure~\ref{fig1}. The
values determined for $\Theta_1$, $\Theta_2$ and $\delta$ depend on the
viewing angle $\Theta_{\rm O}$. The curves for $\Theta_1$ and $\Theta_2$ are
computed for the mean of the fit parameters $a$ and $b$ determined for each
pulse profile individually. The curve for $\delta$ corresponds to the mean
of the parameters $a$, $b$ and $\Delta$. Assuming that $\Theta_{\rm O}$ is
equal to the inclination $i$ of the system, and adopting $i=83^{\circ}$,
then $\Theta_1 \approx 18^{\circ}$, $\Theta_2 \approx 159^{\circ}$ and
$\delta < 5^{\circ}$. The curves determined for the parameters $a$, $b$ and
$\Delta$ for all pulse profiles individually lie within the error bars
displayed at $\Theta_{\rm O}=i=83^{\circ}\pm4^{\circ}$.
}{fig5}
\end{figure}

The remaining ambiguity in the determination of the beam pattern is
indicated by the different units of the lower ($\theta_+$) and the upper
($\theta_-$) x-axis in Figure~\ref{fig4}. However as discussed in
\S~\ref{35d} the study of the evolution of the pulse profile with the 35-day
cycle indicates that the $\theta_+$-solution is presumably the correct
one. The beam pattern of the emission regions has been reconstructed in the
range $66^{\circ}<\theta_+<116^{\circ}$ or $64^{\circ}<\theta_-<114^{\circ}$
(for $\Theta_{\rm O}=83^{\circ}$). The emission regions are unobservable
under values of $\theta_+$ ($\theta_-$) outside this range.

Concerning the decompositions of the short-on pulse profiles, fits of a
quality similar to those found for the main-on are not found. Additionally
the values of the best fit parameters are different from each other and
different from those of the main-on. The same holds true for the pulse
profiles of the observation of the turn-on which unfortunately ended already
when the flux had reached about 2/3 of the maximum flux of this 35-day
cycle, as the lightcurve of the All-Sky-Monitor (ASM) onboard Rossi X-ray
Timing Explorer (RXTE) shows (Wilms 1999, private communication).

\section{Interpretation}
\label{discussion}

The results show that the pulse profiles of Her~X-1 are compatible with the
idea that the beam pattern is symmetric and that a distorted dipole field is
responsible for the asymmetry of the pulse profiles. The analysis does not
permit to discriminate between exact symmetry and a small asymmetry of the
beam pattern. In the case of a small asymmetry, test calculations suggest
that the beam pattern derived above can be regarded as a fair approximation
to the azimuthally averaged beam pattern. Considering the above results a
large asymmetry of the beam pattern seems unlikely. A prominent asymmetry of
the pulse profiles that is primarily due to an asymmetric beam pattern
cannot in general be mimicked by displaced symmetric emission regions,
because one choice of displacement will hardly produce simple and smooth
'false-symmetric constituents' for many different energies and luminosities
with their respective distinct asymmetric pulse shapes. However, the
possibility that the asymmetry of the pulse profiles of Her~X-1 is primarily
due to an asymmetric beam pattern cannot be rigorously excluded. With this
caveat in mind, we will in this section discuss the consequences of the
reconstructed symmetric beam pattern.

\subsection{Beam Pattern}
\label{char}

The beam pattern can also be plotted as a polar diagram with the magnetic
axis ($\theta = 0^{\circ}$) as symmetry axis. This is done in
Figure~\ref{fig6}. It shows the $\theta_+$-solution for the beam pattern in
the energy ranges 6.0~-~8.3~keV (solid line) and 20.0~-~23.0~keV (dashed
line). In the overlap range the mean values of the single-pole contributions
are plotted. Each beam pattern is normalized so that the total power emitted
into the observable solid angle is unity. The $\theta_-$-solution can be
obtained by turning the diagram upside down. No information on the beam
pattern is available in the shaded regions.

\begin{figure}
\picw{3.0in}{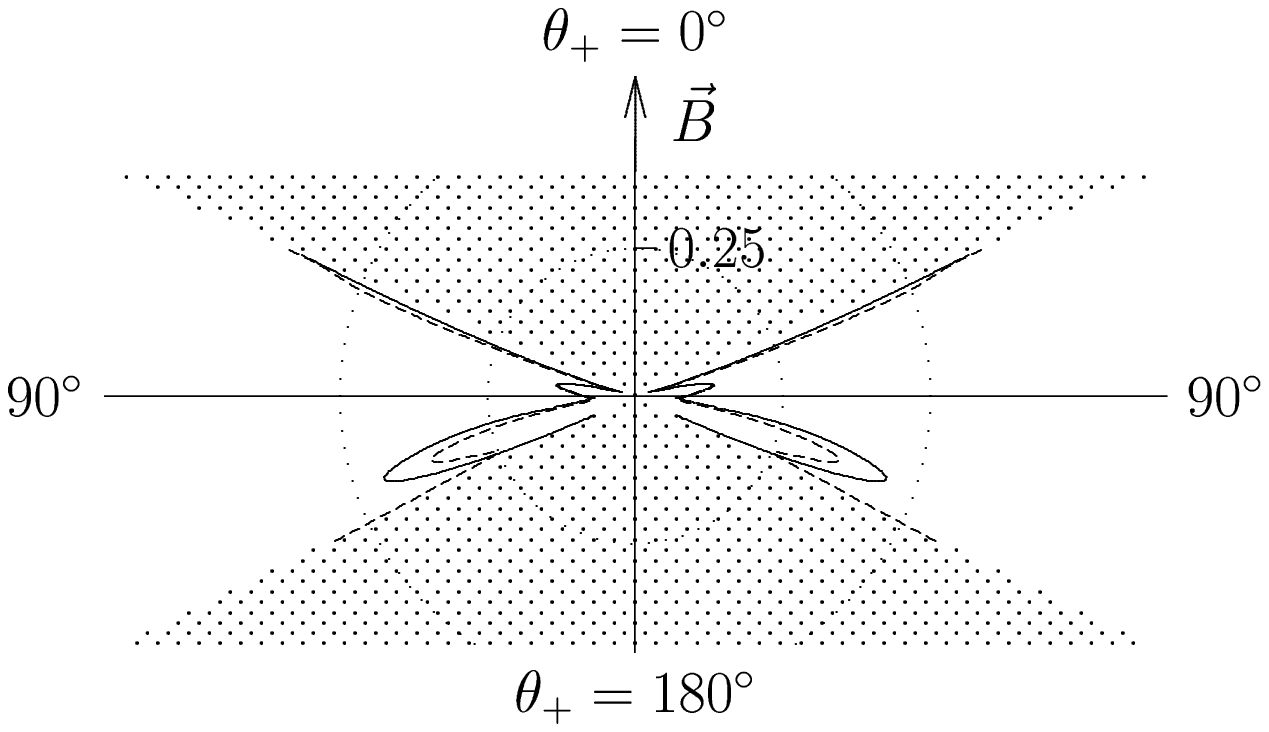}{%
Polar diagram of the beam patterns in the energy ranges 6.0~-~8.3~keV (solid)
and 20.0~-~23.0~keV (dashed).
$\theta_+=0^\circ$ is the direction of the magnetic axis.
}{fig6}
\end{figure}

The visibility of the emission region up to an angle of at least
$116^{\circ}$ is due to a lateral extension of the emission region along the
neutron star surface, emission of radiation from the plasma at a certain
height above the pole and relativistic light deflection near the neutron
star surface. We can get an idea of the effect of light deflection if we
imagine the emission to be originating from a hypothetical point source
located at the pole of the neutron star. With an assumption about the ratio
of the radius of the neutron star $r_{\rm n}$ to its Schwarzschild-radius
$r_{\rm s}$, the asymptotic angle $\theta$ under which the magnetic axis is
seen by the distant observer can be transformed into the intrinsic angle
$\vartheta$ under which the radiation is emitted from the point source (see
Figure~\ref{fig7}). Figure~\ref{fig8} shows how this transformation changes
the asymptotic beam pattern of Figure~\ref{fig6} for $r_{\rm n}/r_{\rm
s}=2.8$. Again the emission pattern is normalized to have an integrated
power of unity. It also illustrates the necessity of taking the effects of
relativistic light deflection into account when modelling the emission
regions, as has been previously pointed out by other authors (e.g. Nollert
et al. 1989).

\begin{figure}
\picw{3.0in}{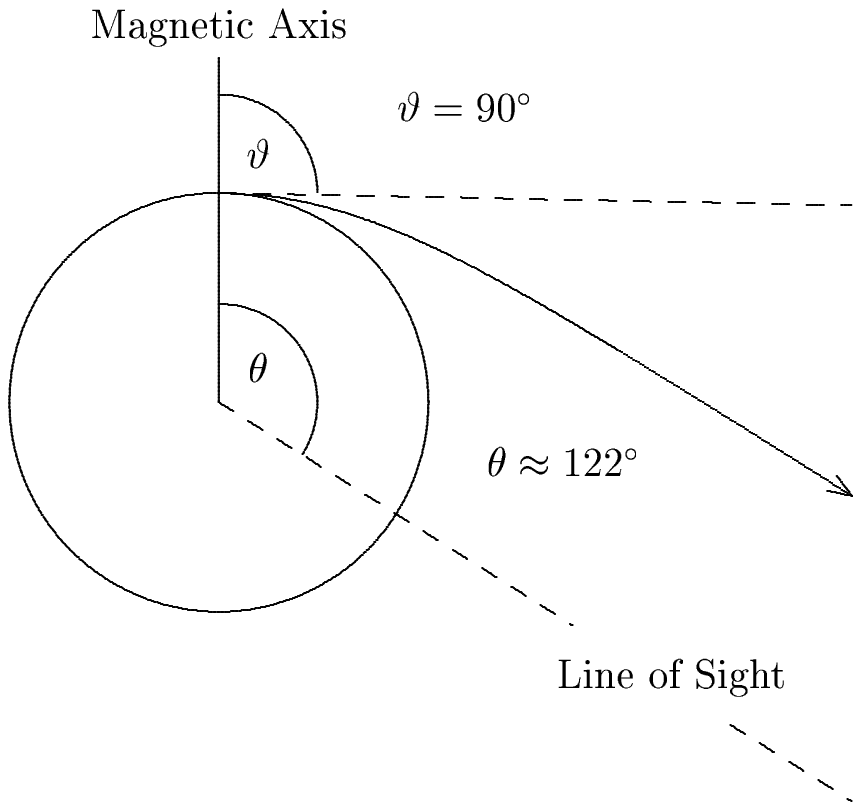}{%
Effect of light deflection on a photon that is emitted at the magnetic pole
on the surface of a neutron star at the local angle $\vartheta=90^{\circ}$
with respect to the magnetic axis. An observer far away from the star
observes the photon at a larger angle $\theta$.  The photon orbit shown in
the figure has been calculated for a neutron star with
$r_{n} / r_{s} = 2.8$ (e.g. $M_{n} = 1.3~M_{\odot}$ and $r_{n} = 10.64~{\rm
km}$).
}{fig7}
\end{figure}
\begin{figure}
\picw{3.0in}{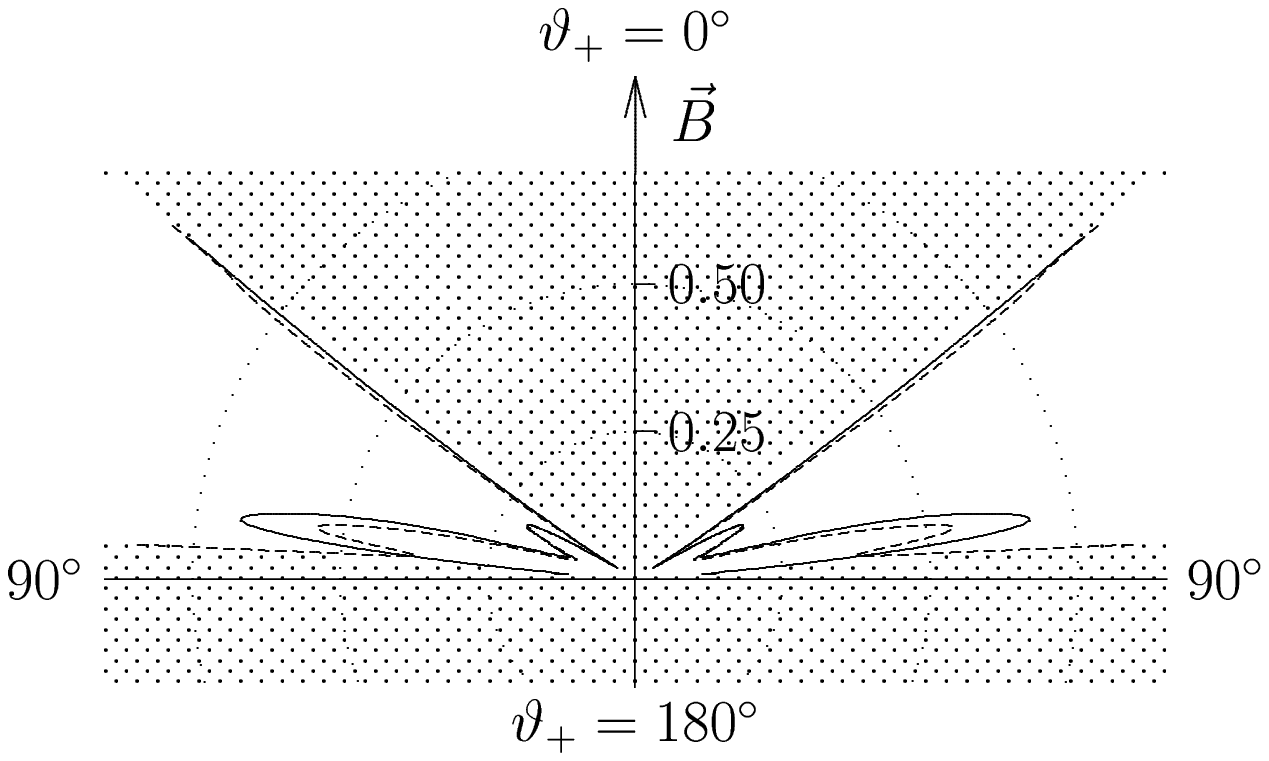}{%
Polar diagram of the intrinsic emission pattern of a hypothetical point
source at the magnetic pole for $r_{\rm n}/r_{\rm s}=2.8$
corresponding to the asymptotic beam patterns of Figure~\ref{fig6}.
}{fig8}
\end{figure}

All beam patterns obtained from the various observations exhibit the same
basic structure and energy dependence. Only the relative sizes of their
substructures differ. The overall structure is quite complex as can be seen
from the representative beam patterns in Figure~\ref{fig6}. It has an
increasing component towards the direction of the magnetic axis. Near the
highest angles of the visible range the flux has a maximum at $\theta_+
\approx 108^\circ$. These major components can be interpreted as a pencil-
and a fan-beam respectively. The relative size of the fan-beam component
decreases with increasing energy. Another relatively small feature occurs at
$\theta_+ \approx 80^\circ$. Above 15~keV the beam pattern has an additional
increasing component at $\theta_+ > 114^\circ$. This feature seems to become
dominant above 28~keV and might therefore be responsible for the observed
widening of the main peak of the pulse profile in this energy range (Soong
1990a, Kuster 1998, private communication). The occurrence of such a feature
in this energy regime indicates a possible relation with electron cyclotron
absorption at about 40~keV, favoured by many authors (e.g. Gruber et
al. 1999). Unfortunately the energy resolution and the statistics of the
data covering the range above 30~keV available to us and suitable for the
analysis was not good enough to give an insight into this property of the
beam pattern.

The beam patterns describe the flux as a function of viewing angle $\theta$
in the various energy ranges. The results can also be plotted as energy
dependent spectra showing the flux depending on energy for various viewing
angles. The left panel in Figure~\ref{fig9} shows 12 beam patterns in the
energy range between 0.92 and 26.0~keV obtained from pulse profiles of an
EXOSAT observation (Kahabka 1987). Since in the pulse profiles the response
of the detectors is not considered, the flux of the beam patterns derived
from the pulse profiles is normalized at the arbitrarily chosen angle
$\theta_+ \approx 90^\circ$ (indicated by an arrow). The angular range is
divided into four sections in which the main features of the beam patterns
are located. The other panels of Figure~\ref{fig9} show spectra at various
viewing angles $\theta_+$. Due to the normalization, the spectrum at the
angle of normalization is just a horizontal line and all other spectra are
relative to that particular one. Each spectrum contains two curves
corresponding to the ME-Argon and the ME-Xenon proportional counters of
EXOSAT. It can be easily seen that the spectra in section III, interpreted
as fan-beam, are very soft compared to the spectra in section I, interpreted
as pencil-beam. The spectra in section IV, interpreted as the high energy
feature above, are even harder than those in section I. It should be pointed
out that there is a great difference between the kind of spectra presented
here and spectra obtained from pulse phase spectroscopy. At a particular
pulse phase the poles are generally seen under different viewing angles and
the spectra from both poles are always superposed. Nevertheless we can
indentify the sections in the beam pattern that are responsible for features
in the pulse profile. Then we compare the spectra in these sections with
spectra at the phases where the corresponding features in the pulse profile
occur. E.g. section III of the beam pattern corresponds to the maxima of the
second single-pole contribution around phases 0.15 and 0.45 (see
Figure~\ref{fig2}) which are responsible for the shoulder in the leading
edge and the secondary maximum in the trailing edge of the peak of the pulse
profile. The hardness ratio at these parts of the pulse profile is
relatively low (see e.g. Deeter et al. 1998) which is consistent with the
soft spectra in section III. On the other hand the hardness ratio at the
peak of the pulse profile is very high corresponding to the sections I and
IV where the spectra are relatively hard as well.

\begin{figure*}\begin{center}
\pic{\textwidth}{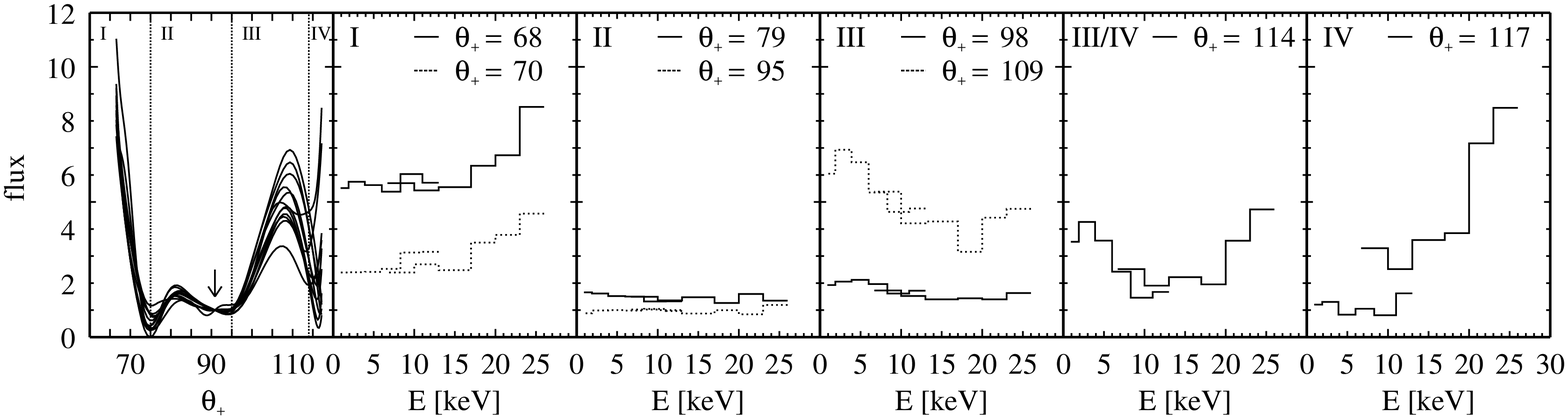}{%
Normalized beam patterns (left) and spectra at various viewing angles
$\theta_+$ obtained from the analysis of energy dependent pulse profiles of
an EXOSAT observation. The normalization point $\theta_+ \approx 90^\circ$
of the beam pattern is indicated by an arrow in the left panel. The range of
the viewing angle has been divided into sections I-IV where the spectra
differ. The two curves of each spectrum correspond to the ME-Argon and the
ME-Xenon proportional counters of EXOSAT. The corresponding section and
viewing angle of each spectrum is given at the top of each panel.
}{fig9}
\end{center}\end{figure*}

Due to the anisotropy of the beam pattern the flux depends on the viewing
angle and therefore on the location of the poles and the inclination of the
system. Then the observed luminosity of the pulsar also depends on the
geometry and the inclination. Since we expect other pulsars to have similar
anisotropic beam patterns but different geometries and inclinations, the
fact that none has a luminosity $L_{\rm x} \gg 10^{38}$ erg/s indicates that
the trend of the flux to increase towards the direction of the magnetic axis
can be expected to reverse at small viewing angles. This would be consistent
with the picture of the radiation escaping into the direction of the
magnetic axis being blocked due to electron cyclotron absorption.

Since the components identified in the energy dependent beam pattern and the
corresponding parts of the pulse profile directly reflect the properties of
the processes of the emission regions, the beam pattern should be further
compared with emission models.

\subsection{Evolution of the Pulse Profile with the 35-day Cycle}
\label{35d}

The evolution of the pulse profile with the 35-day cycle has been studied
intensively by many authors (e.g. Kahabka 1987, \"Ogelman~\&~Tr\"umper 1988,
Soong et al. 1990b, Scott 1993). Deeter et al. (1998) summarize the
observations establishing that the changes in pulse profile throughout the
course of a 35-day cycle are systematic. Several attempts have been made to
explain the change of the pulse shape with the 35-day phase (e.g. Bai 1981,
Tr\"umper et al. 1986, Petterson et al. 1991). In this section we discuss a
scenario in which the column densities along the lines of sight onto the
poles are different due to a partial obscuration of the neutron star by the
inner edge of the accretion disk. This results in a different attenuation of
the polar contributions.

As observed in a two day long continuous monitoring by RXTE, the pulse shape
of Her X-1 does not change significantly during turn-on, whereas the spectra
show strong photoelectric absorption (Kuster et al. 1998). This is in
contrast to the behaviour during the decline of the main-on, when the pulse
shape undergoes systematic changes while no spectral changes are prominent
(Deeter et al. 1998, and references therein). The observations concerning
the spectral behaviour can be explained in terms of a twisted and tilted
accretion disk (Schandl \& Meyer 1994). At turn-on the outer edge of the
warped disk recedes from the line of sight to the neutron star whereas at
the end of the main-on the inner edge sweeps into the line of sight. Since
the obscuring material at the outer edge of the disk is relatively cool
compared to the very dense material at the inner edge, photoelectric
absorption is only present during turn-on. By taking into account the scale
heights of the corresponding parts of the disk, a warped disk profile also
provides a mechanism to explain the different behaviour of the pulse
shape. The density gradient in the obscuring material at the outer edge of
the disk is relatively small. Thus the radiation emerging from both polar
regions experiences the same absorption and the pulse profile does not
change appreciably during the early stages of the main-on. Since on the
other hand, the scale height of the inner edge of the disk is comparable to
the size of the neutron star, the poles become obscured successively towards
the end of the main-on. Therefore the radiation from one pole becomes
attenuated more with respect to the other, leading to changes in the pulse
profile. This situation is schematically illustrated in Figure~\ref{fig10}.
\begin{figure}\begin{center}
\picw{5.5in}{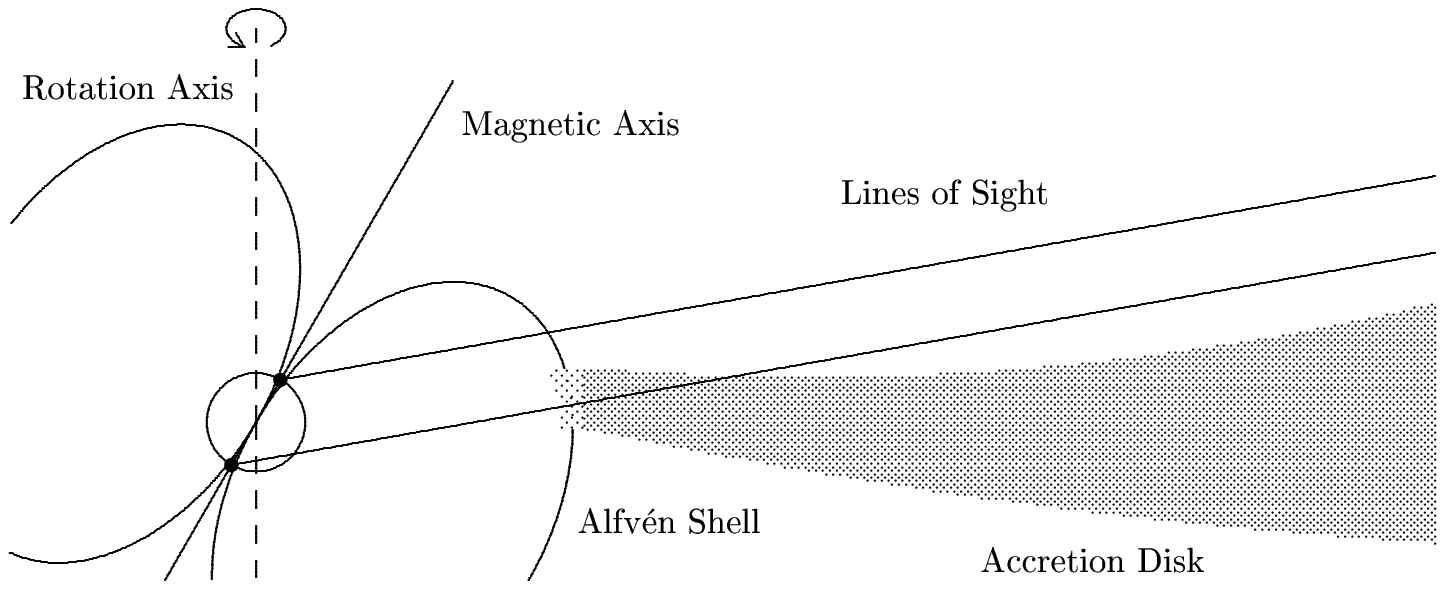}{%
Sketch of the situation during the decay phase of the 35-day cycle. The
innermost part of the warped accretion disk intersects the line of sight
onto the pole located on the far side of the disk with respect to the
observer. Thus the radiation from that pole becomes significantly more
attenuated than the radiation from the other pole. The scale height of the
inner edge of the accretion disk is comparable to the size of the neutron
star.
}{fig10}
\end{center}\end{figure}
The different attenuation of the radiation from the poles is apparent in the
decompositions. Figure~\ref{fig11} shows two pulse profiles of an EXOSAT
observation (Kahabka 1987) during one 35-day cycle at $\Psi_{35}=0.136$ near
maximum intensity (solid) and at $\Psi_{35}=0.234$ during the decay phase of
the main-on state (dotted). The shoulder in the leading edge and the
secondary maximum in the trailing edge of the peak are less prominent in the
decay phase. We find that we can model the pulse profile at the end of the
main-on state with the decompositions found for the pulse profile at maximum
intensity by scaling one component with respect to the other. The pulse
shape plotted with crosses in Figure~\ref{fig11} is reproduced from the
decompositions of the pulse shape at maximum intensity by scaling the second
component by a factor of 0.7 and adding the unscaled first component. It
indeed closely resembles the features of the pulse profile in the decay
phase. We conclude that during the decay phase the neutron star was partly
obscured.

\begin{figure}
\picw{3.in}{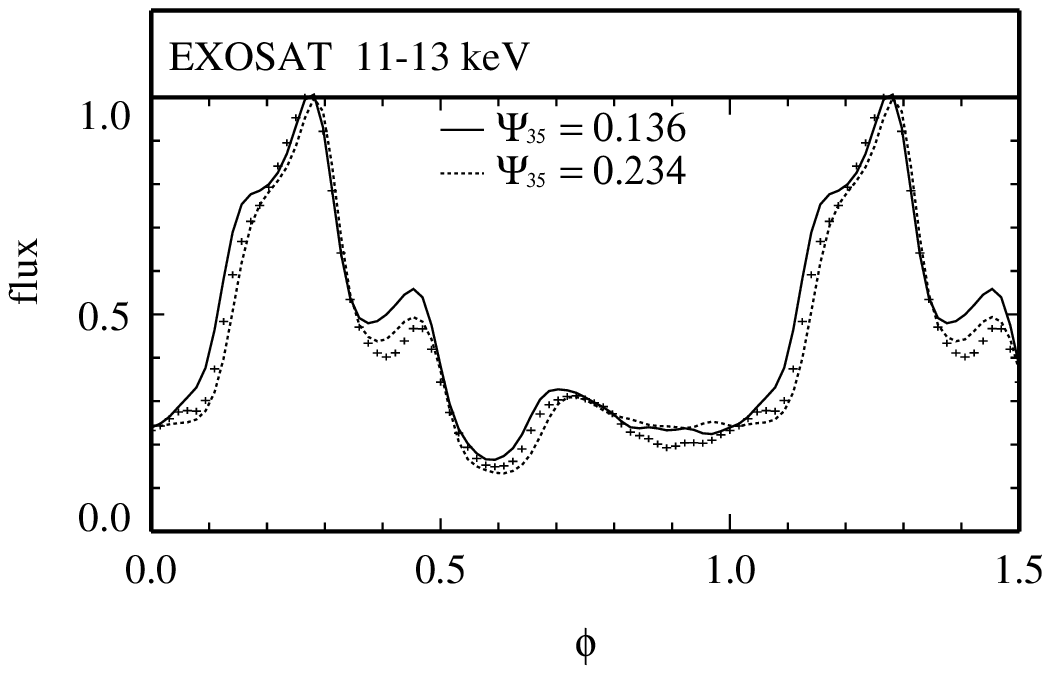}{%
Two pulse profiles of an EXOSAT observation within the same 35-day cycle
during maximum intensity at $\Psi_{35}=0.136$ (solid line) and during the
decay phase at $\Psi_{35}=0.234$ (dotted line). The crosses are a pulse
profile reproduced from the decompositions of the maximum flux pulse profile
with the contribution of the second pole being scaled by a factor of 0.7. It
closely resembles the late phase pulse shape, especially in the main peak.
}{fig11}
\end{figure}

The fact that the second component has to be scaled simply means that the
radiation from the second pole is attenuated more than the radiation from
the first pole. Therefore the second pole must be located on that side of
the neutron star which is on the opposite side of the accretion disk with
respect to the observer. This enables us to decide between the $\theta_-$-
and the $\theta_+$-solution discussed in section~\ref{results}. It follows
that the second component must correspond to the higher values of the
viewing angle $\theta$ and therefore the $\theta_+$-solution must be the
correct one. In a previous analysis of pulse profiles of the X-ray binary
Cen~X-3 (Kraus~et~al.~1996), we have also found unique decompositions and
the beam patterns and their energy dependence are quite similar to those of
Her~X-1. But we were not able to decide between the $\theta_+$- and the
$\theta_-$-solution. However the similarity of the beam patterns suggests
that the $\theta_+$-solution is the correct one for this pulsar, too.

Many authors have noted a narrowing of the main peak during the decay phase
of the 35-day cycle (see Kunz 1996). Different attenuation of the
components obtained in the analysis provides a natural explanation of this
behaviour of the pulse profile.

As the amount of matter along the lines of sight increases, not only
attenuation but also scattering of radiation from other directions into the
line of sight increases. This leads to an increasing fraction of scattered
flux in the pulse profile and the pulsed fraction \footnote{$\mbox{pulsed
fraction}=1-\frac{\mbox{minimum flux of pulse profile}} {\mbox{mean flux of
pulse profile}}$} decreases. Other processes that lead to an increase of
unpulsed flux are reprocessing of the direct beams by the interposed
material or reflection from the disk (McCray et al. 1982). In other words
the pulsed fraction is an indicator of the fraction of radiation that is
coming directly from the polar regions. A pulse profile which contains a
large fraction of scattered flux will have a small pulsed fraction. From
such a pulse profile we can not expect to be able to reconstruct the beam
pattern. Figure~\ref{fig12} shows that indeed the pulse profiles for which
an acceptable fit has been found (denoted by filled symbols) are just those
with a high pulsed fraction. The fact that the analysis of the pulse
profiles of the short-on state has not led to acceptable fits can then be
undestood in terms of the low value of their pulsed fraction. Accounting for
possible attenuation, the components can be scaled in the fit procedure.
This leads for the 15 pulse profiles of three main-on observations with a
flux of about 70\% of the typical maximum flux of the main-on state to a
significant decrease of the deviation $\lambda^2_{\rm red}$
\footnote{
$\lambda^2_{\rm red}=\frac{1}{N-\nu}
\sum_{\rm N}{(f_1({\rm i})-f_2({\rm i}))^2}$, where $\nu$
is the number of fit parameters} between the two curves. These pulse
profiles typically have a medium pulsed fraction.

\begin{figure}
\picw{3.0in}{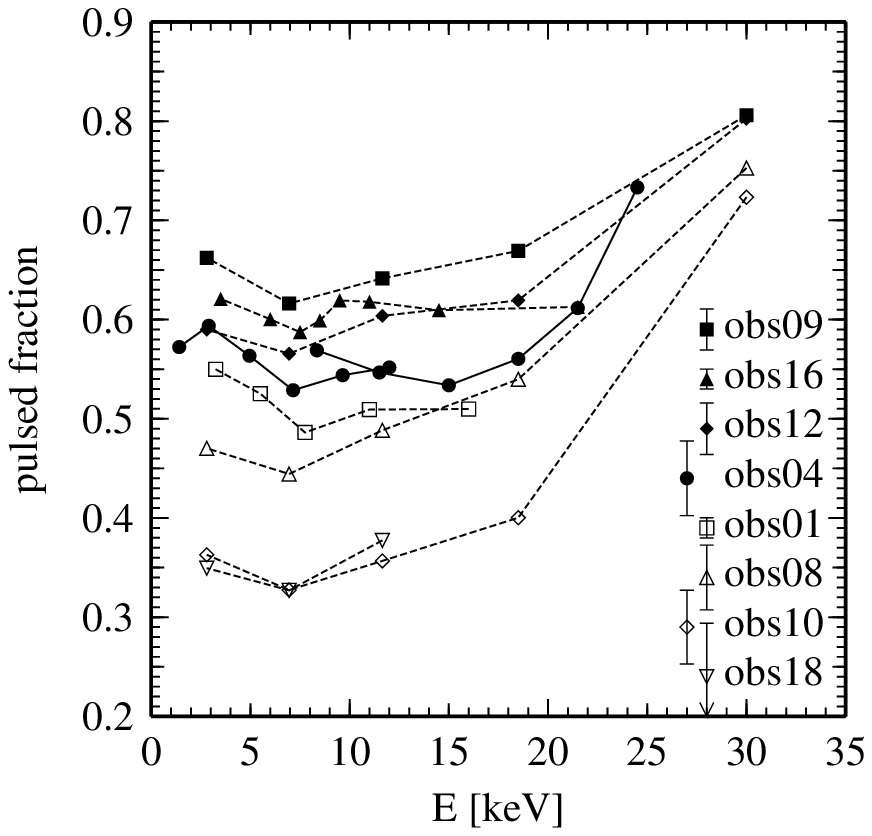}{%
Examples of the pulsed fraction vs energy for different observations. The
filled symbols represent pulse profiles of the main-on state for which the
beam pattern could be reconstructed. The pulse profiles corresponding to the
open symbols are observations during the short-on (obs18), the decline phase
of the main-on (obs10 and obs08) and during the rise-phase of the main-on
(obs01) for which no overlap fit has been found. The typical errors
indicated in the key for the different observations result from error
propagation of the statistical errors of the pulse profiles. The pulsed
fractions of each observation are connected with lines to guide the
eyes. Note the minima of the pulsed fractions near 7~keV which are obviously
due to non-pulsed iron line emission.
}{fig12}
\end{figure}

Observations show that the spectral behaviour at X-ray turn-on is similar
for the short-on and main-on states and that the pulse shape also changes
during short-on (Deeter et al. 1998). This suggests that the configuration
of the disk causing the spectral behaviour and the evolution of the pulse
profile described above in the case of the main-on state is similar during
short-on. Thus the outer part of the disk is responsible for the turn-on of
the short-on state, whereas it ends when the inner edge of the disk passes
into the line of sight.

\acknowledgments

This work has been supported by the Deutsche Forschungsgemeinschaft (DFG).


\end{document}